%% file: paper.tex
\documentclass{nature}
\usepackage{mathcmds}
\usepackage{sansmath}
\usepackage{authblk}
\usepackage{graphicx}
\usepackage{booktabs,tabularx}
\usepackage{subcaption}
\usepackage[hyphens]{url}
\usepackage[colorlinks=true,allcolors=blue]{hyperref}
\usepackage[capitalize]{cleveref}
\usepackage[modulo]{lineno}
\usepackage{todonotes}
\usepackage{threeparttable}
\usepackage{enumitem}
\usepackage{tikz}
\usetikzlibrary{calc, causal, shapes.callouts}

\tikzstyle{predict} = [->, ultra thick, blue, shorten <=3pt, shorten >=2pt, inner sep=2pt, sloped, above, font={\scriptsize\sffamily\itshape}]
\pgfkeys{/tikz/on layer/.code={
    \pgfonlayer{#1}\begingroup
    \aftergroup\endpgfonlayer
    \aftergroup\endgroup
}}
\pgfdeclarelayer{background}
\pgfsetlayers{background,main}

\SetMathAlphabet{\mathsfbf}{sans}{\sansmathencoding}{\sfdefault}{bx}{sl}
\newcommand{\red}[1]{{\color{red}#1}}

\newcommand{\tikzsubfiglabel}[1][south west]{%
    \refstepcounter{subfigure}%
    \node[anchor=#1, inner sep=2pt] at (current bounding box.#1) {\normalsize\bfseries\sffamily\thesubfigure};%
}

\DefineCondParensCommand{\Probs}{P^*}
\DefineCondParensCommand{\Ptrain}{P_\mathrm{tr}}
\DefineCondParensCommand{\ptrain}{p_\mathrm{tr}}
\DefineCondParensCommand{\Ptest}{P_\mathrm{te}}
\DefineCondParensCommand{\ptest}{p_\mathrm{te}}

\newcommand{\figstyle}{\footnotesize\sffamily\sansmath}
\newcommand{\tablestyle}{\footnotesize\sffamily\sansmath}

\crefname{section}{Sec.}{Secs.}

\graphicspath{{fig/}}

\begin{document}

\title{Causality matters in medical imaging}
\author{\sffamily Daniel C.\ Castro\footnote{Corresponding author: \texttt{dc315@imperial.ac.uk}.}, Ian Walker, \& Ben Glocker}

\maketitle

\begin{affiliations}
    \item[] Biomedical Image Analysis Group, Imperial College London, London SW7 2AZ, UK
\end{affiliations}


\input{sections/abstract}


\section*{Introduction}
\input{sections/introduction}

\section*{Causality Matters}
\input{sections/causality_in_mi}
\input{sections/data_scarcity}
\input{sections/data_mismatch}

\section*{Discussion}
\input{sections/conclusion}

\section*{Acknowledgements}
This research has received funding from the European Research Council (ERC) under the European Union's Horizon 2020 research and innovation programme (grant agreement No 757173, project MIRA, ERC-2017-STG), and was partly supported by the CAPES Foundation, Brazil (BEX 1500/2015-05) and the Natural Environment Research Council. The authors also thank Ronald M. Summers for feedback on this manuscript.

\bibliographystyle{naturemag}
\bibliography{cites_long}

\appendix
\section*{Appendix: Background}
\input{sections/background}

\end{document}

%% file: sections/abstract.tex
\begin{abstract}
This article discusses how the language of causality can shed new light on the major challenges in machine learning for medical imaging: 1) data scarcity, which is the limited availability of high-quality annotations, and 2) data mismatch, whereby a trained algorithm may fail to generalize in clinical practice. Looking at these challenges through the lens of causality allows decisions about data collection, annotation procedures, and learning strategies to be made (and scrutinized) more transparently. We discuss how causal relationships between images and annotations can not only have profound effects on the performance of predictive models, but may even dictate which learning strategies should be considered in the first place. For example, we conclude that semi-supervision may be unsuitable for image segmentation---one of the possibly surprising insights from our causal analysis, which is illustrated with representative real-world examples of computer-aided diagnosis (skin lesion classification in dermatology) and radiotherapy (automated contouring of tumours). We highlight that being aware of and accounting for the causal relationships in medical imaging data is important for the safe development of machine learning and essential for regulation and responsible reporting. To facilitate this we provide step-by-step recommendations for future studies.
\end{abstract}

%% file: sections/introduction.tex





Tremendous progress has been achieved in predictive analytics for medical imaging. With the advent of powerful machine-learning approaches such as deep learning, staggering improvements in predictive accuracy have been demonstrated for applications such as computer-aided diagnosis \cite{esteva2017dermatologist} or assisting radiotherapy planning and monitoring of disease progression via automatic contouring of anatomical structures \cite{menze2015multimodal}. However, two of the main obstacles for translating these successes to more applications and into wider clinical practice remain: \emph{data scarcity}, concerning the limited availability of high-quality training data required for building predictive models; and \emph{data mismatch}, whereby a model trained in a lab environment may fail to generalize to real-world clinical data.

Let us illustrate with a hypothetical scenario how these obstacles may arise in practice and pose real threats to the success of research projects. Suppose a team of academic radiologists is excited about the opportunities artificial intelligence seems to offer for their discipline. In a recent study, the clinical team was able to demonstrate the effectiveness of using human interpretation of magnetic resonance imaging (MRI) for diagnosis of prostate cancer, yielding higher sensitivity and specificity than a conventional diagnostic test, as confirmed via ground-truth labels from histopathology. Motivated by these results, the team decides to approach a machine-learning (ML) research lab with the idea of developing a tool for automated, MRI-based diagnosis of prostate cancer.
Because reading MRI requires advanced training and experience, they hope such a system may facilitate widespread adoption of MRI as a novel, accurate, and cost-effective tool for early diagnosis, especially in locations with lower availability of the required human expertise.

The clinicians still have access to their previous study data, and are confident this may be used for ML development. Unfortunately, the sample size is small---there are insufficient pairs of images and diagnosis labels to train a state-of-the-art deep learning image classification method. However, the clinicians have access to large amounts of (unlabelled) routine MRI scans. The ML researchers are hopeful they can additionally leverage this data in a so-called semi-supervised learning strategy. After a pilot phase of method development, the team is planning to evaluate their system in a large multi-centre study.

What are the chances of success for their project, and how could a causal analysis help them to identify potential issues in advance? Regarding the limited availability of annotated data, here the team may be lucky in successfully exploiting the unlabelled data thanks to the anticausal direction between images and confirmed diagnosis labels (as we will discuss later in more detail). However, a major obstacle arises due to a mismatch between the retrospective study data and the prospective multi-centre data due to specific inclusion criteria in the previous study (selection bias), varying patient populations (e.g.\ changes in demographics), and prevalence of disease (e.g.\ due to environmental factors). While researchers are generally aware of the adverse effects of such differences in aspects of the data, they may be unaware that causal reasoning provides tools for laying out any underlying assumptions about the data generating process in a clear and transparent fashion, such that any issues can be more easily identified beforehand and possibly resolved by employing suitable data collection, annotation, and machine-learning strategies.

In this article we discuss how causal considerations in medical imaging can shed new light on the above challenges and help in finding appropriate solutions. In particular, we illustrate how the causal structure of a task can have profound, and sometimes surprising, consequences on the soundness of the employed machine-learning approach and resulting analysis. We highlight that being aware of causal relationships, and related issues such as dataset shift and selection bias, allows for systematic reasoning about what strategies to prefer or to avoid. Here, the language of causal diagrams provides explicit means to specify assumptions, enabling transparent scrutiny of their plausibility and validity \cite{bareinboim2016datafusion}. It is in fact a natural way of defining the relationships between variables of interest, because it reflects the expert's knowledge of the biological and logistical processes involved in the generation and collection of data, and has been successfully applied for building models for decision-making in healthcare, for example \cite{lucas2004bayesnets,cypko2017validation}.
In addition, we provide in the appendix a gentle introduction to the relevant causal-theoretic concepts and notes on creating and interpreting causal diagrams.
We hope our work can serve as a practical guide and inspire new directions for research in medical imaging.


%% file: sections/causality_in_mi.tex

\subsection{Predictive analytics in medical imaging.}

The focus of this article is on predictive modelling: given an image $X$, train a model to predict some given annotation $Y$. Specifically, we wish to estimate the conditional probability distribution ${\Prob{Y \given X}}$ by fitting a statistical model with a suitable objective function. This formulation encompasses a variety of common medical image analysis tasks, such as semantic segmentation (i.e.\ contouring of structures of interest), disease classification, outcome prediction, and many more.

In this context, it is worth clarifying some terminology regarding the data that is used for development and after deployment, in order to avoid confusion of some terms that are sometimes used differently in clinical and machine-learning communities. We refer to an annotated dataset with pairs $(X,Y)$ as the development data, which is used to train and test a predictive model in a lab environment. In ML workflows, the development data is typically split into a training, a validation and a hold-out test set. The training set is used to learn the model parameters (e.g.\ the weights in a convolutional neural network), while the validation set is used during training to monitor the learning progress and avoid overfitting to the training set. The test set is used only after training is completed, in order to quantify the performance of the model on `unseen' data. However, as the development data is often re-used during iterative development cycles, it is well known that information can leak from the test set into training, hence performance reported on the test set can become unrealistic over time \cite{dwork2015holdout}, which poses a major problem for regulators.

Importantly, the assumption that the performance of a trained model on the development test set is representative of the performance on new clinical data after deployment in varying environments is often violated due to differences in data characteristics, as discussed earlier.
Furthermore, contrary to the development test set, the real-world test data after deployment will not come with ground-truth annotations, and performance is thus difficult (or impossible) to assess. It is therefore absolutely critical to be able to clearly formalize and communicate the underlying assumptions regarding the data generating processes in the lab and real-world environments, which in turn can help anticipate and mitigate failure modes of the predictive system.

\subsection{Challenges in medical imaging.}

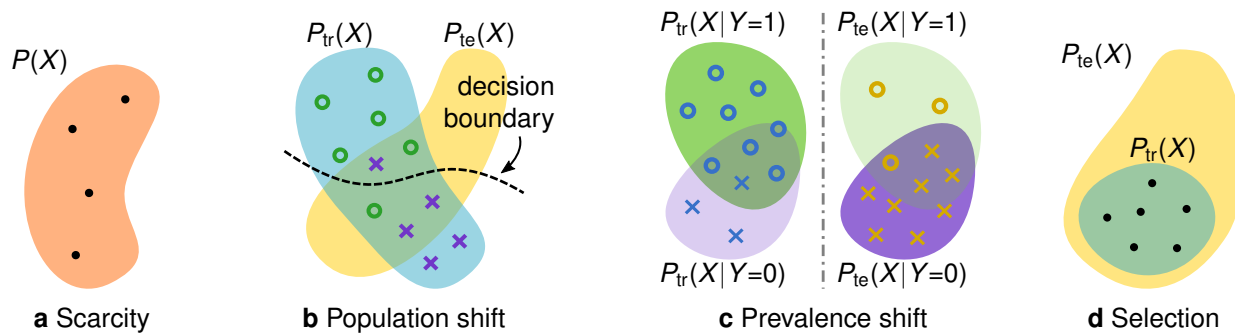
\begin{figure}[tb]
    \centering
    \figstyle
    \let\fbox\relax
    \fbox{\begin{subfigure}[b]{.14\textwidth}
        \centering
		\def\svgwidth{18mm}
        \input{fig/fig1_scarcity.tex}
        \caption{Scarcity}
    \end{subfigure}}
    \hfill
    \fbox{\begin{subfigure}[b]{.25\textwidth}
        \centering
		\def\svgwidth{32mm}
        \input{fig/fig1_covariate.tex}
        \caption{Population shift}
    \end{subfigure}\hspace{1em}}
    \hfill
    \fbox{\begin{subfigure}[b]{.27\textwidth}
        \centering
		\def\svgwidth{42mm}
        \input{fig/fig1_target.tex}
        \caption{Prevalence shift}
    \end{subfigure}}
    \hfill
    \fbox{\begin{subfigure}[b]{.15\textwidth}
        \centering
		\def\svgwidth{24mm}
        \input{fig/fig1_selection.tex}
        \caption{Selection}
    \end{subfigure}}
    \caption{Key challenges in machine learning for medical imaging: \textbf{a} data scarcity and \textbf{b}--\textbf{d} data mismatch. $X$ represents images and $Y$, annotations (e.g.\ diagnosis labels). $\Ptrain$ refers to the distribution of data available for training a predictive model, and $\Ptest$ is the test distribution, i.e.\ data that will be encountered once the model is deployed. Circles and crosses indicate images with different labels (e.g.\ cases vs.\ controls).}
    \label{fig:challenges}
\end{figure}

One of the notorious challenges in medical image analysis is the scarcity of labelled data, in great part due to the high costs of acquiring expert annotations or expensive lab tests, e.g.\ to confirm initial diagnosis. The techniques often used to circumvent this shortage have markedly different properties under the lens of causality. First, we will discuss semi-supervised learning (SSL): the attempt to improve predictive performance by additionally exploiting more abundant unlabelled data. We will follow with an analysis of data augmentation, a powerful paradigm for artificially boosting the amount of labelled data.

In addition, the recurrent issue of mismatch between data distributions, typically between training and test sets or development and deployment environments, tends to hurt the generalizability of learned models. In the generic case with unconstrained disparities, any form of learning from the training set is arguably pointless, as the test-time performance can be arbitrarily poor. Nonetheless, causal reasoning enables us to recognize special situations in which direct generalization is possible, and to devise principled strategies to mitigate estimation biases. In particular, two distinct mechanisms of distributional mismatch can be identified: dataset shift and sample selection bias. Learning about their differences is helpful for diagnosing when such situations arise in practice.

Before diving into the details of these challenges, however---illustrated with cartoon examples in \cref{fig:challenges}---the causal properties of the core predictive task must be analysed. In particular, one must pay close attention to the relationship between the inputs and targets of the model.

\subsection{Causality in medical imaging.}



Given the specification of the input images, $X$, and the prediction targets, $Y$, it is imperative to determine which is the cause and which is the effect. Using the categorization in Ref.~\citenum{scholkopf2012causal}, we wish to establish whether a task is
\begin{itemize}[noitemsep]
    \item \textit{causal:} estimate $\Prob{Y \given X}$, when $X \causes Y$ (predict effect from cause); or
    \item \textit{anticausal:} estimate $\Prob{Y \given X}$, when $Y \causes X$ (predict cause from effect).
\end{itemize}
The answer is crucial to all further causal analysis of the problem, and has a strong impact on the applicability of semi-supervised learning \cite{chapelle2006ssl,scholkopf2013ssl} (discussed later) and on whether generative or discriminative models should be preferred \cite{blobaum2015discriminative}.

Recall the definitions of cause and effect (see Methods): if the annotation could have been different by digitally editing the image beforehand, then one can conclude that the image causes the annotation. For example, manual segmentation masks are drawn over the image by visual inspection and would evidently be influenced by certain pixel changes. On the other hand, a pathology lab result would be unaffected by such manipulations. Images and targets may alternatively be confounded, i.e.\ descend from a common cause. This relationship is often treated similarly to the anticausal case \cite{scholkopf2012causal}.

\begin{table*}[tb]
    \centering
    \tablestyle

    \caption{Causally relevant meta-information}
    \label{tab:meta_info}
    \newlength{\rowskip}
    \setlength{\rowskip}{5pt}
    \begin{tabularx}{\linewidth}{>{\hspace{.5em}}p{37mm}X}
        \toprule
        \bfseries Attribute     & \bfseries Examples \\
        \midrule
        \multicolumn{2}{l}{\itshape Causal direction (predict effect from cause or cause from effect)} \\[\rowskip]
        field of application    & diagnosis / screening / prognosis / exploratory research \\[\rowskip]
        task category           & segmentation / classification / regression / detection \\[\rowskip]
        annotation method       & manual / (semi-)automatic / clinical tests; annotation policy \\[\rowskip]
        nature of annotations   & image-wide label / pixel-wise segmentation / spatial coordinates \\[\rowskip]
        annotation reliability  & image noise, acquisition artefacts, low contrast; user or software errors; signal-to-noise ratio, inter- and intra-observer variability \\
        \midrule
        \multicolumn{2}{l}{\itshape Data mismatch (comparing development vs.\ deployment environments)} \\[\rowskip]
        cohort characteristics  & healthy volunteers / patients; demographics, medical records \\[\rowskip]
        subject selection       & routine / specific condition or treatment / specific age range; quality control \\[\rowskip]
        acquisition conditions  & single- / multi-site; modality; device; vendor; protocol \\[\rowskip]
        train-test split        & random / stratified / balanced \\[\rowskip]
        annotation process      & (see above) \\
        \bottomrule
    \end{tabularx}
\end{table*}

It is generally possible to discern causal structures only when we are aware of the acquired data's background details, as meta-information plays a fundamental role in understanding the data generation and collection processes. Based on a comprehensive ontology of medical imaging meta-information \cite{maierhein2018rankings}, we have compiled in \cref{tab:meta_info} a list of attributes that are meaningful for characterizing the predictive causal direction and detecting signs of dataset mismatch. Let us further illustrate this discussion with two practical examples, depicted in \cref{fig:examples}. Their descriptions will mention some concepts related to dataset mismatch that will be discussed in detail later on.

\subsection{Skin lesion classification example.}
\label{ex:skin_lesions}
Assume a set of dermoscopic images ($X$) is collected along with histopathology diagnosis for melanoma following biopsy ($Y$). Here, $Y$ is a gold-standard proxy for the true presence of skin cancer, and as such can be considered as a \emph{cause} of the visual appearance of the lesion, $X$. This task is therefore \emph{anticausal} (note the arrow directions in \cref{fig:skin_lesion}).

Routine dermoscopic examination of pigmented skin lesions typically results in a `benign', `suspicious', or `malignant' label. Prediction of such labels would instead be \emph{causal}, as they are obtained visually and could be affected if the images were digitally manipulated. Moreover, we know that patients are referred for biopsy only if dermoscopy raises suspicions. As inclusion in this study is case-dependent, a dataset with ground truth biopsy labels suffers from \emph{sample selection bias}, and is thus not representative of the overall distribution of pigmented skin lesions.

\begin{figure}[tb]
    \centering
    \input{fig/examples.tex}
\end{figure}

\subsection{Brain tumour segmentation example.}
\label{ex:segmentation}
Structural brain MRI scans ($X$) are acquired for a cohort of glioma patients, after which a team of radiologists performs manual contouring of each lesion ($Y$). This annotation is done by visual inspection and evidently depends on image content, resolution, and contrast, for example, whereas manually editing the segmentation masks would have no effect on the images. These considerations allow us to conclude this is a case of \emph{causal} prediction ($X \causes Y$).
Here it might also be natural to assume the radiologists were aware of the diagnosis (e.g.\ specific cancer subtype and stage), in which case we could include an additional arrow from `disease' to `segmentation'. This would however not alter the fact that the segmentations are a \emph{consequence} of the images (and diagnoses), thus the task remains \emph{causal}. Regardless, notice how any model trained on this data will be learning to replicate this particular manual annotation process, rather than to predict a `true' underlying anatomical layout.

In addition, suppose our dataset was collected and annotated for research purposes, employing a high-resolution {3\,T} MRI scanner and containing a majority of older patients, and that the trained predictive model is to be deployed for clinical use with conventional {1.5\,T} scanners. This is a clear case of \emph{dataset shift}, firstly because the images are expected to be of different quality (\emph{acquisition shift}). Secondly, because the different age distribution in the target population entails variations in brain size and appearance, and in the prevalences of various types of tumour (\emph{population shift}).%

For the two examples above, establishing the causal direction between images and prediction targets seemed reasonably straightforward. This is not always the case, and arguably in many settings identifying whether the relationship is causal or anti-causal can be non-trivial, particularly if crucial meta-information is missing. Consider the case when prediction targets are extracted from radiology reports. 
At first, one may conclude that the report reflects purely the radiologist's reading of a medical image, hence image causes report. However, their conclusions might be based on additional information---potentially even more important than the findings in the images---such as blood tests or other diagnostic test results. This instance highlights the importance of modelling the full data generating process and of gathering the right information to make an informed decision about the causal relationships underlying the data.

    
    

%% file: fig/fig1_scarcity.tex
\begingroup%
  \makeatletter%
  \providecommand\color[2][]{%
    \errmessage{(Inkscape) Color is used for the text in Inkscape, but the package 'color.sty' is not loaded}%
    \renewcommand\color[2][]{}%
  }%
  \providecommand\transparent[1]{%
    \errmessage{(Inkscape) Transparency is used (non-zero) for the text in Inkscape, but the package 'transparent.sty' is not loaded}%
    \renewcommand\transparent[1]{}%
  }%
  \providecommand\rotatebox[2]{#2}%
  \ifx\svgwidth\undefined%
    \setlength{\unitlength}{84.35170578bp}%
    \ifx\svgscale\undefined%
      \relax%
    \else%
      \setlength{\unitlength}{\unitlength * \real{\svgscale}}%
    \fi%
  \else%
    \setlength{\unitlength}{\svgwidth}%
  \fi%
  \global\let\svgwidth\undefined%
  \global\let\svgscale\undefined%
  \makeatother%
  \begin{picture}(1,1.65658134)%
    \put(0,0){\includegraphics[width=\unitlength,page=1]{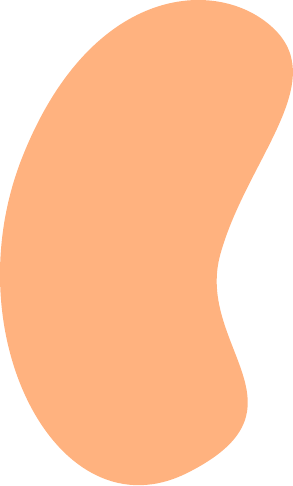}}%
    \put(-0.10363221,1.62191208){\color[rgb]{0,0,0}\makebox(0,0)[lb]{\smash{$\Prob{X}$}}}%
    \put(0,0){\includegraphics[width=\unitlength,page=2]{fig1_scarcity_svg.pdf}}%
  \end{picture}%
\endgroup%

%% file: fig/fig1_covariate.tex
\begingroup%
  \makeatletter%
  \providecommand\color[2][]{%
    \errmessage{(Inkscape) Color is used for the text in Inkscape, but the package 'color.sty' is not loaded}%
    \renewcommand\color[2][]{}%
  }%
  \providecommand\transparent[1]{%
    \errmessage{(Inkscape) Transparency is used (non-zero) for the text in Inkscape, but the package 'transparent.sty' is not loaded}%
    \renewcommand\transparent[1]{}%
  }%
  \providecommand\rotatebox[2]{#2}%
  \ifx\svgwidth\undefined%
    \setlength{\unitlength}{142.61274269bp}%
    \ifx\svgscale\undefined%
      \relax%
    \else%
      \setlength{\unitlength}{\unitlength * \real{\svgscale}}%
    \fi%
  \else%
    \setlength{\unitlength}{\svgwidth}%
  \fi%
  \global\let\svgwidth\undefined%
  \global\let\svgscale\undefined%
  \makeatother%
  \begin{picture}(1,1.08070738)%
    \put(0,0){\includegraphics[width=\unitlength,page=1]{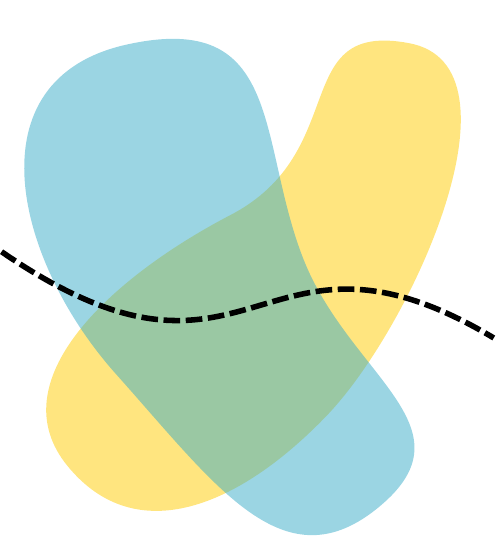}}%
    \put(0.08885064,1.02743264){\color[rgb]{0,0,0}\makebox(0,0)[lb]{\smash{$\Ptrain{X}$}}}%
    \put(0.65901993,1.02743264){\color[rgb]{0,0,0}\makebox(0,0)[lb]{\smash{$\Ptest{X}$}}}%
    \put(0.62620876,0.67444718){\color[rgb]{0,0,0}\makebox(0,0)[lb]{\smash{\parbox[b]{.5\unitlength}{\raggedleft decision boundary}}}}%
    \put(0,0){\includegraphics[width=\unitlength,page=2]{fig1_covariate_svg.pdf}}%
  \end{picture}%
\endgroup%

%% file: fig/fig1_target.tex
\begingroup%
  \makeatletter%
  \providecommand\color[2][]{%
    \errmessage{(Inkscape) Color is used for the text in Inkscape, but the package 'color.sty' is not loaded}%
    \renewcommand\color[2][]{}%
  }%
  \providecommand\transparent[1]{%
    \errmessage{(Inkscape) Transparency is used (non-zero) for the text in Inkscape, but the package 'transparent.sty' is not loaded}%
    \renewcommand\transparent[1]{}%
  }%
  \providecommand\rotatebox[2]{#2}%
  \ifx\svgwidth\undefined%
    \setlength{\unitlength}{176.54027856bp}%
    \ifx\svgscale\undefined%
      \relax%
    \else%
      \setlength{\unitlength}{\unitlength * \real{\svgscale}}%
    \fi%
  \else%
    \setlength{\unitlength}{\svgwidth}%
  \fi%
  \global\let\svgwidth\undefined%
  \global\let\svgscale\undefined%
  \makeatother%
  \begin{picture}(1,0.87376937)%
    \put(-.02,0.81539674){\color[rgb]{0,0,0}\makebox(0,0)[lb]{\smash{$\Ptrain{X|Y{=}1}$}}}%
    \put(-.02,0.02106115){\color[rgb]{0,0,0}\makebox(0,0)[lb]{\smash{$\Ptrain{X|Y{=}0}$}}}%
    \put(0,0){\includegraphics[width=\unitlength,page=1]{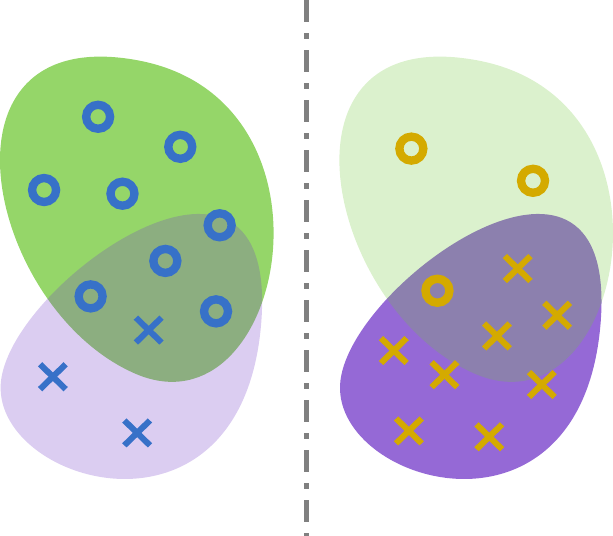}}%
    \put(0.54,0.81539674){\color[rgb]{0,0,0}\makebox(0,0)[lb]{\smash{$\Ptest{X|Y{=}1}$}}}%
    \put(0.54,0.02106115){\color[rgb]{0,0,0}\makebox(0,0)[lb]{\smash{$\Ptest{X|Y{=}0}$}}}%
  \end{picture}%
\endgroup%

%% file: fig/fig1_selection.tex
\begingroup%
  \makeatletter%
  \providecommand\color[2][]{%
    \errmessage{(Inkscape) Color is used for the text in Inkscape, but the package 'color.sty' is not loaded}%
    \renewcommand\color[2][]{}%
  }%
  \providecommand\transparent[1]{%
    \errmessage{(Inkscape) Transparency is used (non-zero) for the text in Inkscape, but the package 'transparent.sty' is not loaded}%
    \renewcommand\transparent[1]{}%
  }%
  \providecommand\rotatebox[2]{#2}%
  \ifx\svgwidth\undefined%
    \setlength{\unitlength}{116.1178647bp}%
    \ifx\svgscale\undefined%
      \relax%
    \else%
      \setlength{\unitlength}{\unitlength * \real{\svgscale}}%
    \fi%
  \else%
    \setlength{\unitlength}{\svgwidth}%
  \fi%
  \global\let\svgwidth\undefined%
  \global\let\svgscale\undefined%
  \makeatother%
  \begin{picture}(1,1.3268755)%
    \put(0,0){\includegraphics[width=\unitlength,page=1]{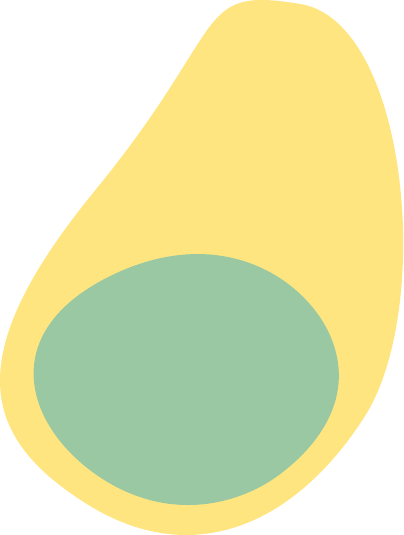}}%
    \put(0.35709528,0.73127298){\color[rgb]{0,0,0}\makebox(0,0)[lb]{\smash{$\Ptrain{X}$}}}%
    \put(-.02,1.25324794){\color[rgb]{0,0,0}\makebox(0,0)[lb]{\smash{$\Ptest{X}$}}}%
    \put(0,0){\includegraphics[width=\unitlength,page=2]{fig1_selection_svg.pdf}}%
  \end{picture}%
\endgroup%

%% file: fig/examples.tex
\tikzset{x=1.4cm, y=1.2cm}

\begin{subfigure}[c]{.46\textwidth}
    \centering
    \begin{tikzpicture}
        \node[obs, label={left:\strut disease}]    at (0,1)  (y) {};
        \node[obs, label={below:image}]            at (1,1)  (x) {};
        \node[obs, label={right:\strut suspicion}] at (2,1)  (z) {};
        \node[sel, label={right:\strut selection}] at (2,0)  (s) {};
        
        \draw (y) edge (x);
        \draw (x) edge (z);
        \draw (z) edge (s);

        \draw[predict, transform canvas={yshift=2mm}] (x) -- node {\itshape predict} (y);

        \tikzsubfiglabel[south west]{}
        \label{fig:skin_lesion}
    \end{tikzpicture}
\end{subfigure}\hfill%
\begin{subfigure}[c]{.5\textwidth}
    \centering
    \begin{tikzpicture}
        \node[obs, label={right:\strut segmentation}]       at (2,0)  (y) {};
        \node[obs, label={below:image}]                     at (1,0)  (x) {};
        \node[obs, label={left:\strut disease}]             at (0,0)  (z) {};
        \node[dom, label={right:\strut train / test domain}]  at (.5,1) (d) {};
        
        \draw (x) edge (y);
        \draw (z) edge (x);
        \draw (d) edge (x) edge (z);
        
        \draw[predict, transform canvas={yshift=2mm}] (x) -- node {\itshape predict} (y);

        \tikzsubfiglabel[north west]{}
        \label{fig:brain_tumour}
    \end{tikzpicture}
\end{subfigure}

\caption{Causal diagrams for medical imaging examples. \textbf{a}~Skin lesion classification. \textbf{b}~Brain tumour segmentation. Filled circular nodes represent measured variables, double circular nodes denote sample selection indicators, and squares are used for sample domain indicators. Here we additionally highlight the direction of the predictive task.}
\label{fig:examples}

%% file: sections/data_scarcity.tex


\subsection{Tackling data scarcity via semi-supervision.}\label{sec:ssl}

Semi-supervised learning (SSL) aims to leverage readily available unlabelled data in the hope of producing a better predictive model than is possible using only the scarce annotated data. 
Given this ambitious goal, it is perhaps unsurprising that strong requirements need to be met. Namely, the distribution of inputs needs to carry relevant information about the prediction task---otherwise it would be pointless to collect additional unlabelled data. This idea is typically articulated in terms of specific assumptions about the data which can be intuitively summarised as follows \cite{chapelle2006ssl}: similar inputs (images in our case) are likely to have similar labels and will naturally group into clusters with high density in the input feature space. Lower density regions in that space in-between clusters are assumed to be ideal candidates for fitting decision boundaries of predictive models. In this context, considering large amounts of unlabelled data together with the scarce labelled data may reveal such low density regions and may lead to better decision boundaries than using labelled data alone.

Note how this idea insinuates an interplay between the distribution of inputs, $\Prob{X}$, and the label conditional, $\Prob{Y \given X}$. Now recall that, by independence of cause and mechanism (see Methods), if the prediction task is causal ($X \causes Y$), then $\Prob{X}$ is uninformative with respect to ${\Prob{Y \given X}}$, and SSL is theoretically futile in this case \cite{chapelle2006ssl,scholkopf2013ssl}. Since typical semantic segmentation tasks are \emph{causal}, as illustrated in our brain tumour example, there is likely very little hope that semantic segmentation can fundamentally benefit from unlabelled data, which may relate to recent concerns raised in the literature \cite{oliver2018evaluation}. Conversely, if $Y \causes X$ as for skin lesions, then these distributions \emph{may} be dependent, and semi-supervision has a chance of success \cite{scholkopf2013ssl}. We conjecture that, in practice, anticausal problems are more likely than causal ones to comply with the SSL assumptions outlined above, as observed e.g.\ among the datasets analysed in Ref.~\citenum{blobaum2015discriminative}.


That is not to say that SSL is completely useless for causal tasks, as there can be practical algorithmic benefits. Under certain conditions, unlabelled data can be shown to have a regularizing effect, potentially boosting the accuracy of an imperfect model by lowering its variance \cite{cozman2006risks}, and may reduce the amount of labelled data required to achieve a given performance level \cite{singh2008unlabeled,balcan2006pac}. Further work is needed to empirically validate these gains in causal and anticausal scenarios.

A recent comprehensive empirical study \cite{oliver2018evaluation} reported that properly tuned purely supervised models and models pre-trained on related labelled datasets (i.e.\ transfer learning) are often competitive with or outperform their semi-supervised counterparts. It also demonstrated that SSL can hurt classification performance under target shift (discussed later as \emph{prevalence shift}) between labelled and unlabelled sets. This suggests that practitioners willing to apply SSL should be cautious of potential target distribution mismatch between labelled and unlabelled sets---e.g.\ unequal proportions of cases and controls or presence of different pathologies.



\subsection{Tackling data scarcity via data augmentation.}\label{sec:data_augmentation}

Data augmentation refers to the practice of systematically applying random, controlled perturbations to the data in order to produce additional plausible data points. This now ubiquitous technique aims to improve the robustness of trained models to realistic variations one expects to find in the test environment, and has met tremendous practical success across a wide variety of tasks. Notably, we can distinguish between augmentations encouraging \emph{invariance} and \emph{equivariance}.

Many tasks require predictions to be insensitive to certain types of variation. Examples include image intensity augmentations, such as histogram manipulations or addition of noise, and spatial augmentations (e.g.\ affine or elastic transformations) for image-level tasks (e.g.\ regression or classification, as in the skin lesion example). As these augmentations apply uniformly to all inputs $x$ without changing the targets $y$, their benefits stem from a refined understanding of the conditional $\Prob{X \given Y}$, while contributing no new information about $\Prob{Y}$.

For other tasks, such as segmentation or localization, predictions must change similarly to the inputs, e.g.\ a spatial transformation applied to an image $x$---such as mirroring, affine or elastic deformations---should be likewise applied to the target $y$ (e.g.\ spatial coordinates or segmentation masks, as in the brain tumour example). Information is gained about the joint distribution via its shared spatial structure, related to e.g.\ anatomy and acquisition conditions.

In contrast with SSL, data augmentation produces additional $(x, y)$ pairs, thereby providing more information about the \emph{joint} distribution, $\Prob{X,Y}$. Its compound effect on the joint $\Prob{X,Y}$ rather than only on the marginal $\Prob{X}$ corroborates its suitability for both causal and anticausal tasks, without the theoretical impediments of semi-supervised learning for causal prediction.

An emerging line of research aims to exploit unlabelled data for \emph{learning} realistic transformations for data augmentation \cite{zhao2019augmentation,chaitanya2019augmentation}. This direction has the potential to deliver the promises of semi-supervised learning while improving over the reliable framework of standard data augmentation.

%% file: sections/data_mismatch.tex


\subsection{Data mismatch due to dataset shift.}\label{sec:dataset_shift}

Dataset shift is any situation in which the training and test data distributions disagree due to exogenous factors, e.g.\ dissimilar cohorts or inconsistent acquisition processes. As before, let $X$ be the input images and $Y$ be the prediction targets. We use an indicator variable $D$ for whether we are considering 
the training ($\Ptrain{X,Y}$) or the test domain ($\Ptest{X,Y}$):
\begin{equation}
    \Ptrain{X,Y} \defeq \Prob[D=0]{X,Y} \quad\text{and}\quad
    \Ptest{X,Y} \defeq \Prob[D=1]{X,Y} \,.
\end{equation}
For simplicity, in the following exposition we will refer only to disparities between training and test domains. This definition can however extend to differences between the development datasets (training and test data) and the target population (after deployment), when the latter is not well represented by the variability in the test data.

Moreover, when analysing dataset shift, it is helpful to conceptualize an additional variable $Z$, representing the unobserved physical reality of the subject's anatomy. We then interpret the acquired images $X$ as imperfect and potentially domain-dependent measurements of $Z$, i.e.\ ${Z\causes X}$. 

Switching between domains may produce variations in the conditional relationships between $X$, $Y$, and $Z$ or in some of their marginal distributions. Based on the predictive causal direction and on which factors of the joint distribution change or are invariant across domains, dataset shift can be classified into a variety of familiar configurations. Here we formulate the concepts of `population shift', `annotation shift', `prevalence shift', `manifestation shift', and `acquisition shift'. These terms correspond roughly to particular dataset shift scenarios studied in general machine-learning literature, namely `covariate shift', `concept shift', `target shift', `conditional shift', and `domain shift', respectively \cite{quinonerocandela2009shift}. However, we believe it is beneficial to propose specific nomenclature that is more vividly suggestive of the phenomena encountered in medical imaging. By also explicitly accounting for the unobserved anatomy, the proposed characterization is more specific and enables distinguishing cases that would otherwise be conflated, such as population or manifestation shift versus acquisition shift. The basic structures are summarized in \cref{fig:dataset_shift} in the form of selection diagrams (causal diagrams augmented with domain indicators) \cite{bareinboim2016datafusion}, and some examples are listed in \cref{tab:dataset_shift}. We hope this may empower researchers in our field to more clearly communicate dataset shift issues and to more easily assess the applicability of various solutions.


\begin{figure*}[tb]
    \centering
    \input{fig/dataset_shift.tex}
    \label{fig:dataset_shift}
\end{figure*}

\begin{table*}[tb]
    \centering
    \tablestyle

    \newcommand{\addrowskip}{\rule{0pt}{3ex}}
    \newcommand{\colskip}{\hskip 1em}
    \newcolumntype{L}[1]{>{\raggedright\let\newline\\\arraybackslash\hspace{0pt}}p{#1}}
    \renewcommand{\given}{\vert}
    \renewcommand{\Prob}[2][]{P_{#1}(#2)}
    \renewcommand{\prob}[2][]{p_{#1}(#2)}

    \begin{threeparttable}
    \caption{Types of dataset shift}
    \label{tab:dataset_shift}
    \begin{tabular}{l@{\colskip}l@{\colskip}l@{\colskip}l}
        \toprule
        \bfseries Type      & \bfseries Direction   & \bfseries Change    & \bfseries Examples of differences \\
        \midrule
        Population shift    & causal                & $\Prob[D]{Z}$
            & ages, sexes, diets, habits, ethnicities, genetics \\
        \addrowskip%
        Annotation shift    & causal                & $\Prob[D]{Y\given X}$
            & annotation policy, annotator experience \\
        \addrowskip%
        Prevalence shift    & anticausal            & $\Prob[D]{Y}$
            & case-control balance, target selection \\
        \addrowskip%
        Manifestation shift & anticausal            & $\Prob[D]{Z\given Y}$
            & anatomical manifestation of the target disease or trait \\
        \addrowskip%
        Acquisition shift   & either                & $\Prob[D]{X\given Z}$
            & scanner, resolution, contrast, modality, protocol \\
        \bottomrule
    \end{tabular}
    \end{threeparttable}
\end{table*}

For causal prediction, we name \emph{population shift} the case wherein only intrinsic characteristics (e.g.\ demographics) of the populations under study differ, i.e.\ $\Ptrain{Z}\neq\Ptest{Z}$. Fortunately, this case is \emph{directly transportable}, i.e.\ a predictor estimated in one domain is equally valid in the other \cite{pearl2014validity}. An underfitted model (`too simple') may however introduce spurious dependencies, for which importance reweighting with $\ptest{x}/\ptrain{x}$ is a common mitigation strategy \cite{storkey2009transfer,zhang2015multisource}. Clearly, learning in this scenario makes sense only if the variability in the training data covers the support of the test distribution \cite{quinonerocandela2009shift}---in other words, there are no guarantees about extrapolation performance to modes of variation that are missing from the training environment.


Under \emph{prevalence shift} (for anticausal tasks), the differences between datasets relate to class balance: ${\Ptrain{Y}\neq\Ptest{Y}}$. This can arise for example from different predispositions in the training and test populations, or from variations in environmental factors. If the test class distribution $\Ptest{Y}$ is known \textit{a priori} (e.g.\ from an epidemiological study), generative models may reuse the estimated appearance model $\Ptrain{X \given Y}$ ($=\Ptest{X \given Y}$) in Bayes' rule, and, for discriminative models, instances can be weighted by $\ptest{y}/\ptrain{y}$ to correct the bias in estimating the training loss. Alternatively, more elaborate solutions based on the marginal $\Ptest{X}$ are possible \cite{storkey2009transfer,zhang2013target}.

Cases of \emph{annotation shift} involve changes in class definitions, i.e.\ the same datum would tend to be labelled differently in each domain (${\Ptrain{Y \given X}\neq\Ptest{Y \given X}}$). For example, it is not implausible that some health centres involved in an international project could be operating slightly distinct annotation policies or grading scales, or employing annotators with varying levels of expertise (e.g.\ senior radiologists vs.\ trainees). Without explicit assumptions on the mechanism behind such changes, models trained to predict $\Ptrain{Y \given X}$ evidently cannot be expected to perform sensibly in the test environment, and no clear solution can be devised \cite{morenotorres2012unifying}. A tedious and time-consuming calibration of labels or (partial) re-annotation may be required to correct for annotation shift.

Another challenging scenario is that of \emph{manifestation shift}, under which the way anticausal prediction targets (e.g.\ disease status) physically manifest in the anatomy changes between domains. In other words, ${\Ptrain{Z\given Y}\neq\Ptest{Z\given Y}}$. As with annotation shift, this cannot be corrected without strong parametric assumptions on the nature of these differences.

We lastly discuss \emph{acquisition shift}, resulting from the use of different scanners or imaging protocols, which is one of the most notorious and well-studied sources of dataset shift in medical imaging \cite{glocker2019multisite}. Typical pipelines for alleviating this issue involve spatial alignment (normally via rigid registration and resampling to a common resolution) and intensity normalization. In addition, the increasingly active research area of domain adaptation investigates data harmonization by means of more complex transformations, such as extracting domain-invariant representations \cite{ganin2016domain,kamnitsas2017unsupervised} or translating between imaging modalities \cite{frangi2018simulation} (e.g.\ synthesizing MRI volumes from CT scans \cite{huo2019synsegnet}).


\red{




}

\subsection{Data mismatch due to sample selection bias.}\label{sec:sample_selection}

A fundamentally different process that also results in systematic data mismatch is sample selection. It is defined as the scenario wherein the training and test cohorts come from the same population, though each training sample is measured ($S=1$) or rejected ($S=0$) according to some selection process $S$ that may be subject-dependent:
\begin{equation}
    \Ptrain{X,Y} \defeq \Prob{X,Y \given S=1} \quad\text{and}\quad
    \Ptest{X,Y} \defeq \Prob{X,Y} \,.
\end{equation}
The main difference to standard dataset shift is the \emph{data-dependent} selection mechanism (\cref{tab:sample_selection}), as opposed to external \emph{causes} of distributional changes (\cref{fig:dataset_shift}). In other words, the indicator variables in sample selection concern alterations in the data-gathering process rather than in the data-generating process \cite{zhang2015multisource}.



\begin{figure*}[tb]
    \centering
    \input{fig/sample_selection.tex}
    \label{fig:sample_selection}
\end{figure*}

\begin{table*}[tb]
    \centering
    \tablestyle

    \newcommand{\addrowskip}{\rule{0pt}{3ex}}
    \newcommand{\colskip}{\hskip 1em}

    \captionof{table}{Types of sample selection}
    \label{tab:sample_selection}
    \begin{tabularx}{\textwidth}{l@{\colskip}l@{\colskip}>{\raggedright}X@{\colskip}l}
        \toprule
        \bfseries Type    & \bfseries Causation & \bfseries Examples of selection processes  & \bfseries Resulting bias \\
        \midrule
        Random      & none
            & uniform subsampling, randomized trial
            & none \\
        \addrowskip%
        Image       & $X{\causes}S$
            & visual phenotype selection (e.g.\ anatomical traits, lesions)
            & population shift \\
        \addrowskip%
        &   & image quality control (QC; e.g.\ noise, low contrast, artefacts)
            & acquisition shift \\
        \addrowskip%
        Target      & $Y{\causes}S$
            & hospital admission, filtering by disease, annotation QC, learning strategies (e.g.\ class balancing, patch selection)
            & prevalence shift \\
        \addrowskip%
        Joint       & $X{\causes}S{\causedby}Y$
            & combination of the above (e.g.\ curated benchmark dataset)
            & spurious assoc. \\
        \bottomrule
    \end{tabularx}
\end{table*}

Completely random selection simply corresponds to uniform subsampling, i.e.\ when the training data can be assumed to faithfully represent the target population (${\Ptrain{X,Y} \equiv \Ptest{X,Y}}$). Since the analysis will incur no bias, the selection variable $S$ can safely be ignored. We conjecture this will rarely be the case in practice, as preferential data collection is generally unavoidable without explicit safeguards and careful experimental design.


Selection can be affected by the appearance of each image in two different manners. We can select subjects based on anatomical features---viewing the image $X$ as a proxy for the anatomy $Z$---which has similar implications to population shift. Alternatively, selection criteria may relate to image quality (e.g.\ excluding scans with noise, poor contrast, or artefacts), which is akin to acquisition shift \cite{morenotorres2012unifying}. If selection is purely image-based ($X \causes S$), we may exploit the conditional independence $S \indep Y \given X$, which implies that the predictive relation is directly \emph{recoverable} \cite{bareinboim2014selection}, i.e.\ $\Ptest{Y \given X} \equiv \Ptrain{Y \given X}$. In a learning scenario, however, the objective function would still be biased, and methods for mitigating the corresponding cases of dataset shift can be employed.



When selection is solely target-dependent ($Y \causes S$), we have $\Ptest{X \given Y} \equiv \Ptrain{X \given Y}$, and it can be treated as prevalence shift. This will typically result from factors like hospital admission, recruitment or selection criteria in clinical trials, or annotation quality control. Notably, machine-learning practitioners should be wary that it can also arise as a side-effect of certain training strategies, such as class re-balancing or image patch selection for segmentation (e.g.\ picking only patches containing lesion pixels).

Sample selection can additionally introduce spurious associations when the selection variable $S$ is a common effect of $X$ and $Y$ (or of causes of $X$ and $Y$): \emph{implicitly conditioning} on $S$ unblocks an undesired causal path between $X$ and $Y$ (see Methods). This is the classic situation called selection bias \cite{hernan2004structural} (cf.\ Berkson's paradox \cite{pearl2009causality}), and recovery is more difficult without assumptions on the exact selection mechanism. In general, it requires controlling for additional variables to eliminate the indirect influence of $X$ on $Y$ via conditioning on the collider $S$ \cite{bareinboim2014selection,bareinboim2016datafusion}.





%% file: fig/dataset_shift.tex
{
\sansmath
\tikzset{x=1cm, y=.7cm}
\renewcommand{\given}{\vert}
\renewcommand{\Prob}[2][]{P_{#1}(#2)}
\begin{subfigure}{.32\textwidth}
    \centering
    \begin{tikzpicture}
        \node[unobs,label={left:$Z$}]   at (0,0) (z) {};
        \node[obs,  label={90:$X$}]     at (1,0) (x) {};
        \node[obs,  label={right:$Y$}]  at (2,0) (y) {};
        \node[dom,  label={$D$}]    at (0,1) (d) {};
        \draw (z) edge (x);
        \draw (x) edge (y);
        \draw (d) edge (z);
    \end{tikzpicture}
    \caption{\centering Population shift: ${\mathsfbf{\Prob[D]{Z}}\Prob{X\given Z}\Prob{Y\given X}}$}
    \label{fig:covariate_shift}
\end{subfigure} \hfill
\begin{subfigure}{.34\textwidth}
    \centering
    \begin{tikzpicture}
        \node[unobs,label={left:$Z$}]   at (0,0) (z) {};
        \node[obs,  label={45:$X$}]     at (1,0) (x) {};
        \node[obs,  label={right:$Y$}]  at (2,0) (y) {};
        \node[dom,  label={$D$}]        at (1,1) (d) {};
        \draw (z) edge (x);
        \draw (x) edge (y);
        \draw (d) edge (x);
    \end{tikzpicture}
    \caption{\centering (Causal) Acquisition shift: ${\Prob{Z}\mathsfbf{\Prob[D]{X\given Z}}\Prob{Y\given X}}$}
    \label{fig:concept_shift}
\end{subfigure} \hfill
\begin{subfigure}{.32\textwidth}
    \centering
    \begin{tikzpicture}
        \node[unobs,label={left:$Z$}]   at (0,0) (z) {};
        \node[obs,  label={90:$X$}]     at (1,0) (x) {};
        \node[obs,  label={right:$Y$}]  at (2,0) (y) {};
        \node[dom,  label={$D$}]        at (2,1) (d) {};
        \draw (z) edge (x);
        \draw (x) edge (y);
        \draw (d) edge (y);
    \end{tikzpicture}
    \caption{\centering Annotation shift: ${\Prob{Z}\Prob{X\given Z}\mathsfbf{\Prob[D]{Y\given X}}}$}
    \label{fig:conditional_shift}
\end{subfigure} \\[2ex]
\begin{subfigure}{.32\textwidth}
    \centering
    \begin{tikzpicture}
        \node[obs,  label={left:$X$}]   at (0,0) (x) {};
        \node[unobs,label={90:$Z$}]     at (1,0) (z) {};
        \node[obs,  label={right:$Y$}]  at (2,0) (y) {};
        \node[dom,  label={$D$}]        at (2,1) (d) {};
        \draw (y) edge (z);
        \draw (z) edge (x);
        \draw (d) edge (y);
    \end{tikzpicture}
    \caption{\centering Prevalence shift: ${\Prob{X\given Z}\Prob{Z\given Y}\mathsfbf{\Prob[D]{Y}}}$}
    \label{fig:target_shift}
\end{subfigure} \hfill
\begin{subfigure}{.32\textwidth}
    \centering
    \begin{tikzpicture}
        \node[obs,  label={left:$X$}]   at (0,0) (x) {};
        \node[unobs,label={135:$Z$}]    at (1,0) (z) {};
        \node[obs,  label={right:$Y$}]  at (2,0) (y) {};
        \node[dom,  label={$D$}]        at (1,1) (d) {};
        \draw (y) edge (z);
        \draw (z) edge (x);
        \draw (d) edge (z);
    \end{tikzpicture}
    \caption{\centering Manifestation shift: ${\Prob{X\given Z}\mathsfbf{\Prob[D]{Z\given Y}}\Prob{Y}}$}
\end{subfigure} \hfill
\begin{subfigure}{.32\textwidth}
    \centering
    \begin{tikzpicture}
        \node[obs,  label={left:$X$}]   at (0,0) (x) {};
        \node[unobs,label={90:$Z$}]     at (1,0) (z) {};
        \node[obs,  label={right:$Y$}]  at (2,0) (y) {};
        \node[dom,  label={$D$}]        at (0,1) (d) {};
        \draw (y) edge (z);
        \draw (z) edge (x);
        \draw (d) edge (x);
    \end{tikzpicture}
    \caption{\centering (Anticausal) Acquisition shift: ${\mathsfbf{\Prob[D]{X\given Z}}\Prob{Z\given Y}\Prob{Y}}$}
\end{subfigure}
}

\caption{\sansmath Selection diagrams for \textbf{a}--\textbf{c} causal and \textbf{d}--\textbf{f} anticausal dataset shift scenarios, with corresponding factorizations of the joint distribution $\Prob[D]{X,Y,Z}$. $X$ is the acquired image; $Y$, the prediction target; $Z$, the unobserved true anatomy; and $D$, the domain indicator (0: `train', 1: `test'). An unfilled node means the variable is unmeasured.}


%% file: fig/sample_selection.tex
\sansmath
\begin{subfigure}[t]{.23\textwidth}
    \centering
    \begin{tikzpicture}
        \node[obs, label={left:$X$}]    at (0,1) (x) {};
        \node[obs, label={right:$Y$}]   at (2,1) (y) {};
        \node[sel, label={below:$S$}]   at (1,0) (s) {};
        \draw (y) edge[undir] (x);
        \tikzsubfiglabel[south west];
    \end{tikzpicture}
\end{subfigure} \hfill
\begin{subfigure}[t]{.23\textwidth}
    \centering
    \begin{tikzpicture}
        \node[obs, label={left:$X$}]    at (0,1) (x) {};
        \node[obs, label={right:$Y$}]   at (2,1) (y) {};
        \node[sel, label={below:$S$}]   at (1,0) (s) {};
        \draw (y) edge[undir] (x);
        \draw (x) edge (s);
        \tikzsubfiglabel[south west];
    \end{tikzpicture}
\end{subfigure} \hfill
\begin{subfigure}[t]{.23\textwidth}
    \centering
    \begin{tikzpicture}
        \node[obs, label={left:$X$}]    at (0,1) (x) {};
        \node[obs, label={right:$Y$}]   at (2,1) (y) {};
        \node[sel, label={below:$S$}]   at (1,0) (s) {};
        \draw (y) edge[undir] (x);
        \draw (y) edge (s);
        \tikzsubfiglabel[south west];
    \end{tikzpicture}
\end{subfigure} \hfill
\begin{subfigure}[t]{.23\textwidth}
    \centering
    \begin{tikzpicture}
        \node[obs, label={left:$X$}]    at (0,1) (x) {};
        \node[obs, label={right:$Y$}]   at (2,1) (y) {};
        \node[sel, label={below:$S$}]   at (1,0) (s) {};
        \draw (y) edge[undir] (x);
        \draw (x) edge (s);
        \draw (y) edge (s);
        \tikzsubfiglabel[south west];
    \end{tikzpicture}
\end{subfigure}

\caption{Causal diagrams for different sample selection scenarios. \textbf{a} Random; \textbf{b} image-dependent; \textbf{c} target-dependent; \textbf{d} jointly dependent. $S=1$ indicates an observed sample, and plain edges represent either direction.}

%% file: sections/conclusion.tex
\label{sec:conclusion}%
This paper provides a fresh perspective on key challenges in machine learning for medical imaging using the powerful framework of causal reasoning. Not only do our causal considerations shed new light on the vital issues of data scarcity and data mismatch in a unifying approach, but the presented analysis can hopefully serve as a guide to develop new solutions.
Perhaps surprisingly, causal theory also suggests that the common task of semantic segmentation may not fundamentally benefit from unannotated images via semi-supervision. This possibly controversial conclusion may prompt empirical research into validating the feasibility and practical limitations of this approach.

\begin{figure}
    \centering
    \figstyle
    \input{fig/generic_diagram.tex}
\end{figure}


Other advanced topics could be worth exploring in future work for causally expressing more subtle facets of predictive modelling workflows. In particular, one recurring topic in epidemiology and sociology that is relevant to our imaging context is \emph{measurement bias} \cite{hernan2009measurement,shahar2009causal}. This is the study of properties of proxy variables, which stand in for true variables of interest that are difficult or impossible to measure directly. Of particular note are the cases wherein proxies are additionally affected by other variables (`differential'), or when measurement errors for separate proxies are correlated (`dependent') \cite{lash2014practices}. Measurement bias was explored here for the case of acquisition shift (images as proxies for anatomy, affected by the domain), and similar considerations could extend to other variables, e.g.\ patient records or pathology results.

A further pertinent topic is that of \emph{missingness}. Whereas sample selection refers to the observability of full records, missingness concerns partial measurements---i.e.\ when some subjects may be missing observations of some variables. This is the context of semi-supervised learning, for example, as target labels are observed only for a subset of the data points. The classical characterization distinguishes whether data is \emph{missing completely at random}, \emph{missing at random}, or \emph{missing not at random}, when the missingness of a measurement is independent of any of the variables of interest, dependent on observed variables, or dependent on the missing values, respectively \cite{rubin1976missing}. Causal diagrams again prove instrumental in identifying such structural assumptions about missingness mechanisms \cite{daniel2012missing,mohan2013missing}.

Finally, we highlight that our contribution is only the first step towards incorporating causality in medical image analysis. Here we introduce to this community purely the language of causal reasoning, hoping this will facilitate novel research directions exploiting causality theory to its full extent. Specifically, the endeavours of \emph{causal inference} and \emph{causal discovery} are so far largely unexplored in medical imaging.
In this context, they could lead to the discovery of new imaging biomarkers and to exciting new applications such as personalized counterfactual predictions (`What if a patient were not a smoker?').
Large population imaging studies such as the UK Biobank \cite{miller2016ukbiobank,conroy2019ukbiobank} can greatly empower this kind of research, as they offer unique opportunities for extracting the relevant patterns of variation from sheer observational data.

\begin{table}[tb]
    \centering
    \tablestyle

    \caption{Step-by-step recommendations}
    \label{tab:steps}

    \begin{tabularx}{\textwidth}{X}
    \toprule%
    \begin{minipage}[c]{.99\linewidth}
    \begin{enumerate}[topsep=0pt, partopsep=0pt, leftmargin=6mm, after=\strut]
        \item Gather meta-information about the data collection and annotation processes to reconstruct the full story of the dataset (\cref{tab:meta_info}).
        \item Establish the predictive causal direction: does  the image cause the prediction target or vice versa?
        \item Identify any evidence of mismatch between datasets (\cref{tab:dataset_shift}):
        \begin{itemize}[nosep, leftmargin=6mm]
            \item If causal (image $\causes$ target): population shift, annotation shift
            \item If anticausal (target $\causes$ image): prevalence shift, manifestation shift
        \end{itemize}
        \item Verify what types of differences in acquisition are expected, if any.
        \item Determine whether the data collection was biased with respect to the population of interest, and whether selection was based on the images, the targets, or both (\cref{tab:sample_selection}).
        \item Draw the full causal diagram including postulated direction, shifts, and selections.
    \end{enumerate}
    \end{minipage} \\
    \bottomrule
    \end{tabularx}
\end{table}

Beside enabling new research directions, incorporation of causal reasoning in medical image analysis aligns with a growing awareness among stakeholders of the need for responsible reporting in this field.
There have been increasing efforts from regulatory bodies---such as the US Food and Drug Administration \cite{FDA2019,parikh2019regulation}, the UK's Department of Health and Social Care \cite{UKDHSC2018}, National Institute for Health and Care Excellence \cite{NICE2019}, and NHSX \cite{NHSX2019}, and even the World Health Organization \cite{wiegand2019who}---to outline best practices for the safe development and monitoring of AI-enabled medical technologies \cite{lancetdighealth2019guidelines}. Guidelines for designing and reporting traditional clinical trials are now also being specialized for AI-based interventions \cite{liu2019reporting}.
This has been accompanied by a recent surge in discussion among the medical community about the opportunities and, crucially, the risks of deploying such tools in clinical practice \cite{ghassemi2019guidance,prevedello2019challenges,langlotz2019roadmap,wiens2019roadmap,vancalster2019predictive,shah2018recalibrating,shah2019useful,AOMRC2019}.
Most of the apprehension revolves around the \emph{external validity} of these predictive models, i.e.\ their generalizability beyond the development environment, in terms of e.g.\ robustness to dataset shift \cite{ghassemi2019guidance,prevedello2019challenges} and selection bias \cite{hoffman2013misuse,ghassemi2019guidance}, as discussed herein. Other important concerns involve data inaccuracy, inconsistency, and availability \cite{hoffman2013misuse,ghassemi2019guidance,shah2018recalibrating,prevedello2019challenges,langlotz2019roadmap}, and alignment of the model training objective with the target clinical setting \cite{prevedello2019challenges,wiens2019roadmap,shah2018recalibrating,shah2019useful}.
In a similar yet complementary vein to the notable TRIPOD guidelines \cite{collins2015tripod,collins2019tripodml}, our work ties precisely into this context of encouraging transparent reporting of predictive analytics in healthcare.

This debate also relates to parallel initiatives from within the machine learning community, in specific in the emerging field of fairness, accountability, and transparency (FAT). Scholars in FAT have proposed checklist-style guidelines for reporting datasets \cite{gebru2018datasheets} and models \cite{mitchell2019cards}, for example, and have been investigating sources of failure for machine-learning models, among which is poor reporting \cite{saria2019tutorial}. Interestingly, the same formalism of causal reasoning explored here was also shown to be especially well-suited for expressing and addressing issues of unfairness (e.g.\ social biases) \cite{chiappa2019causal} and dataset shift \cite{subbaswamy2019preventing} in other contexts.

Overall, the goal of this article has been to introduce to the medical imaging community the language of causal diagrams, and to demonstrate how it can illuminate common issues in predictive modelling.
While causal reasoning by itself may not solve any of the data scarcity or mismatch problems, it provides a clear and precise framework for expressing assumptions about the data.
Presenting such assumptions transparently in the form of causal diagrams makes them immediately recognizable by other researchers, and therefore easier to be confirmed or disputed.
The real challenge lies in identifying these very assumptions, as they can often be unclear or ambiguous.

To facilitate this task, we offer in \cref{tab:steps} a step-by-step summary of our recommendations, and \cref{fig:generic_diagram} presents a generic `scaffold' diagram from which most typical workflows can be adapted. Readers may then refer to the other tables for help in identifying the components of their own diagram for the problem at hand. We believe that this exercise of building the full causal story of a dataset will encourage analysts to consider potential underlying biases more thoroughly, and that it may, like the TRIPOD checklist, lead to `more comprehensive understanding, conduct, and analysis of prediction model studies' \cite{collins2019tripodml}.

%% file: fig/generic_diagram.tex
\definecolor{popshiftcolor}{HTML}{d95f02}
\definecolor{acqshiftcolor}{HTML}{7570b3}
\definecolor{annshiftcolor}{HTML}{e7298a}
\definecolor{selectcolor}{HTML}{1b9e77}
\tikzstyle{highlight} = [draw=#1, line width=8pt, opacity=.4, line cap=round, line join=round, on layer=background]
\tikzstyle{callout} = [fill opacity=.3, text opacity=1, rectangle callout, callout absolute pointer={#1}, font={\sffamily\scriptsize}]

\begin{tikzpicture}[x=1.8cm, y=1.56cm]
    \node[obs, label={right:\strut annotation ($Y_4$)}]          at (3.5,2)    (seg) {};
    \node[obs, label={[yshift=-1.5ex]right:image ($X$)}]               at (2,2)    (img) {};
    \node[unobs, label={left:\strut disease}]   at (0,2)    (dis) {};
    \node[dom, label={above:train / test ($D$)}]      at (2,4)    (dom) {};
    \node[obs, label={[align=right]170:patient\\characteristics\\\vspace{-\baselineskip}($Y_2$)\rule{1ex}{0pt}}] at (.5,3)   (age) {};
    \node[obs, label={[align=left, yshift=-1.5mm]right:acquisition\\conditions}]    at (2,3)  (scn) {};
    \node[obs, label={left:\strut diagnosis ($Y_1$)}]   at (0,1)    (dx)  {};
    \node[obs, label={[align=left]right:annotation\\conditions}] 
                                                at (3.5,3)  (exp) {};
    \node[unobs, label={[xshift=-3mm]below:anatomy ($Z$)}]         at (1,2)    (ana) {};
    \node[sel, label={below:selection ($S$)}]         at (1,.3)    (sel) {};
    \node[obs, label={right:\strut referral ($Y_3$)}]  at (2,1)    (sus) {};
    
    \edge{dom}{exp};
    \chain{dom, age, dis, ana, img};
    \chain{dis, ana, img, seg};
    \edge{age}{ana};
    \edge{exp}{seg};
    \edge{dis}{dx};
    \edge{dx, sus}{sel};
    \edge{img}{sus};
    \chain{dom, scn, img, seg};
    
    \draw[highlight=popshiftcolor] (dom.center) -- (age.center) -- (dis.center) -- (img.center) (age.center) -- (ana.center);
    \draw[highlight=annshiftcolor] (dom.center) -- (exp.center) -- (seg.center);
    \draw[highlight=acqshiftcolor] (dom.center) -- (img.center);
    \draw[highlight=selectcolor] (dis.center) -- (dx.center) -- (sel.center) -- (sus.center) -- (img.center);
    
    \node[callout=($(dom)!.7!(age)$), fill=popshiftcolor, anchor=east, align=center] at (1.1,4.2) {population shift \textcolor{red}{(causal)}\\prevalence shift \textcolor{blue}{(anticausal)}};
    \node[callout=($(dom)!.8!(exp)$), fill=annshiftcolor, anchor=west] at (3.4,3.7) {annotation shift};
    \node[callout=($(dom)!.6!(scn)$), fill=acqshiftcolor, anchor=west] at (2.9,4.3) {acquisition shift};
    \node[callout=($(dx)!.3!(sel)$), fill=selectcolor, anchor=east] at (.1,.2) {sample selection};
    
    \draw[predict, red, transform canvas={yshift=2mm}] (img) -- node {\itshape predict?} (seg);
    \draw[predict] (img) to[bend left=10] node[below] {\itshape predict?} (dx);
    \draw[predict] (img) to[bend right=15] node {\itshape predict?} (age);
    \draw[predict, red, transform canvas={xshift=-2mm}] (img) -- node[below, pos=.6] {\itshape predict?} (sus);
\end{tikzpicture}
\caption{A `scaffold' causal diagram summarizing typical medical imaging workflows.
We believe most practical cases can be adapted from this generic structure by removing or adding elements.
Here are represented a variety of possible prediction targets (marked $Y_1$--$Y_4$): some anticausal ($Y_1$, $Y_2$) and others, causal ($Y_3$, $Y_4$). `Annotation' here refers to any image-derived data, such as lesion descriptions, regions of interest, spatial landmark coordinates, or segmentation maps.
Note that annotators will often be aware of the patients' records and diagnoses, in which cases there could be additional arrows from $Y_1$ or $Y_2$ towards $Y_4$.}
\label{fig:generic_diagram}

%% file: sections/background.tex

\subsection{Fundamentals of causal reasoning.}\label{sec:background}


Learning tasks can be broadly divided into three categories based on the causal information used: i) \emph{prediction}, in which observed data are used to infer values of unobserved variables, e.g.\ image classification; ii) \emph{interventions}, where investigators study the impact of forcing a variable to attain a certain value, e.g.\ randomized controlled trials (RCTs) for drug testing; and iii) \emph{counterfactual analysis}, wherein observed data combined with a structural causal model are used to answer questions of the form, `What would have happened if individual $I$ had received treatment $T$ instead?'. While most are familiar with causal inference in the context of RCTs or scientific experiments, causal information is vital even in certain purely predictive tasks, as we discussed in the context of medical imaging. 

Causation can be formalized as follows: a variable $A$ is generally said to be a \emph{direct cause} of variable $B$, written $A \causes B$, if forcing $A$ to different values changes the likelihood of $B$, all else held constant \cite{pearl2009causality}. In the language of \emph{structural causal models}, this means that $B$ has some functional dependence on $A$ (potentially also on other factors and on independent noise), called its \emph{mechanism} \cite{peters2017elements}. Crucially, when $A \causes B$, the causal model posits that the distribution of the cause, $P(A)$, does not inform or influence the conditional $P(B \given A)$, a principle known as \emph{independence of cause and mechanism} \cite{daniusis2010inferring,peters2017elements}.%

Consider the example wherein a radiologist makes a decision, $B$, for referral to further clinical testing (e.g.\ needle biopsy) based on any suspicious findings in the patient's medical scan, $A$. Given an image, the distribution over possible decisions is the conditional $\Prob{B \given A}$. If the appearance of the scan changes, this referral distribution---reflecting the radiologist's judgement---changes as well.
On the other hand, the mechanism that translates from a finding of a suspicious pattern in the scan $A$ to a referral decision $B$ is independent of how likely any individual scan is to appear in the real world, $\Prob{A}$.
This is further justified as such mechanism may typically be formed by rules from radiology guidelines.
Here, the cause of the referral decision is clearly the medical scan, as altering the decision would not affect the scan's appearance.

In the above example, the correct graphical model would be $A \causes B$, as resolved via domain knowledge. If presented only with observational data of medical images and referrals, however, from a purely statistical perspective one would find it difficult to identify whether $A \causes B$ or $B \causes A$. It may still be possible to identify which is the correct relationship if the gathered data were the result of two experiments, respectively manipulating $A$ or $B$. 
Determining the presence and direction of causal relationships from data is the realm of \emph{causal discovery}, which is an extremely challenging and active field of research but is beyond the scope of this article.

\subsection{Causal graphical models.}
\newcommand{\parents}[1]{\operatorname{pa}(#1)}
When multiple variables are involved, causal assumptions can be expressed as a simple directed acyclic graph (DAG; no loops, at most one edge between any pair of nodes), whose nodes represent variables of interest and edges between them indicate postulated direct causal influences. Such a causal graphical model, referred to as a \emph{causal diagram}, embodies the causal Markov assumption (or local Markov): every node is statistically independent of its non-effects (non-descendants), given its direct causes (parents). Therefore, the joint probability distribution over all variables $V_i$ on the graph can be factorized as a product of independent conditional mechanisms \cite{peters2017elements,greenland2011diagrams}:
\begin{equation}
    \Prob{V_1, V_2, \dots, V_N} = \prod_{i=1}^N \Prob{V_i \given \parents{V_i}} \,.
\end{equation}
where $\parents{V_i}$ denotes the set of parents of variable $V_i$, i.e.\ the nodes with arrows pointing toward $V_i$.

For those familiar with Bayesian networks, it appears that there is nothing new.
However, Bayesian networks only encode conditional independence relationships, and they are thus not unique for a given observational distribution \cite{pearl2009causality}.
In fact, although causal arguments often guide the construction of such models, any alignment between arrows in Bayesian networks and causality is merely coincidental. 
In particular, causal models differ from Bayesian networks in that, beside representing a valid factorization of the joint probability distribution, they enable reasoning about \emph{interventions} \cite{pearl2009causality}. In causal graphs, the values for each node are assumed to be determined via \emph{independent mechanisms} (cf.\ independence of cause and mechanism) given their direct causes. An intervention is defined as any forced change to the value or distribution of a node, regardless of its direct causes, and results in a modified graph wherein this node is disconnected from its parents, though crucially all other mechanisms are unaffected.
This can also be thought of as replacing the mechanism generating a variable by a function independent of its former direct causes (e.g.\ a constant).
Incidentally, this is the principle behind randomized controlled trials: a treatment is assigned at random (an intervention on the `treatment' variable), isolating its direct effect on the outcome by eliminating the influence of confounding factors (i.e.\ cutting the edges from common causes of treatment and outcome).
Note that considering interventions on image and referral decision is also what allowed us to determine the causal direction in the example above.

\newcommand{\inlinetikz}[2][]{\tikz[baseline=-\the\dimexpr\fontdimen22\textfont2\relax,#1]{#2}}
\subsection{Building a causal diagram.}
The first step in constructing a causal model for a given system is to elicit the relevant variables to represent, which may be observed or not. These ought to be well defined: they should unambiguously correspond to real or postulated entities of the system, and separate variables must not have overlapping meanings \cite{swanson2013public}. In the medical imaging context, variables normally correspond to the collected data elements, such as images, meta-information fields, labels, patient records, etc. Not all important variables need to be concrete and measurable, however. Other relevant abstract concepts can be instantiated if they help in describing complex processes: e.g.\ `annotation policy`, `patient's health status', `proprietary image post-processing pipeline'.

Secondly, the causal links between the defined variables must be determined. While each added arrow between two nodes in the graph corresponds to assuming causation, it is important to consider that the \emph{absence} of an arrow also encodes a strong assumption. Namely, that there is no direct causal effect---any marginal association between those variables is fully explained via mediator variables or common causes (see below).
In addition, the granularity of `direct effects' is only relative to the chosen level of abstraction \cite{greenland2011diagrams}. One may wish to detail the complete chain of effects between two causally linked variables, or represent them by a single arrow (e.g.\ $A \causes B_1 \causes B_2 \causes C$ vs.\ $A \causes C$).


In what is called a \emph{selection diagram} \cite{pearl2014validity}, one also includes special indicator variables that identify the `domain' or `environment', e.g.\ training vs.\ testing or a certain hospital in a multi-site study. Their direct causal effects (outgoing arrows) represent the specific mechanisms through which one assumes the observed populations differ, whereas the absence of a link from a domain selector to a variable implies the latter's mechanism is invariant across environments \cite{pearl2014validity}. Domain indicators should normally be represented by root nodes in the diagram, with no incoming edges, as they embody \emph{exogenous} changes to the data distributions. A causal diagram may additionally be augmented with selection variables, when the dataset is subject to preferential subsampling from the population (e.g.\ inclusion criteria for a clinical trial). The incoming arrows to such a node represent the various selection criteria (deliberate or otherwise) that impacted the collection of the dataset of interest.

Finally, note that this construction is an iterative process. Once a full version of the diagram is written, one must verify that the assumptions implied by the graph match the domain knowledge (see following notes on interpretation), and corrections should be made as needed. Further, recall the diagram's intent as a communication tool when choosing its level of abstraction, as there is often a tradeoff to be made between accuracy and clarity: the graph should be sufficiently detailed not to omit relevant variables and pathways, though no more complex than necessary \cite{swanson2013public}.




\subsection{Interpreting causal diagrams.}

Given a causal diagram, it is possible to read from it the assumed direct and indirect causal links and conditional independences. In particular, let us analyse some canonical relationships between three variables. If $A$ affects $C$ indirectly through its impact on $B$ ($A \causes B \causes C$), we say $B$ is a \emph{mediator} and $A$ is an \emph{indirect cause} of $C$. Here, $B$ completely screens off the effect of $A$ on $C$, meaning $A \indep C \given B$ (read `$A$ is conditionally independent of $C$, given $B$').
Alternatively, assume $B$ is a common cause of $A$ and $C$: $A \causedby B \causes C$. In this case, $B$ is known as a \emph{confounder}, producing an association between $A$ and $C$, thus $A \notindep C$ (read `$A$ is not independent of $C$'). However, controlling for $B$ makes them independent: $A \indep C \given B$.
Finally, consider the case wherein $B$ is a common effect of $A$ and $C$ ($A \causes B \causedby C$), called a \emph{collider}. Unlike the two situations above, this configuration implies $A$ and $C$ are independent \textit{a priori}. On the other hand, conditioning on $B$ introduces an association between $A$ and $C$, as they may now `explain away' the effect of each other on the observed outcome, $B$ (i.e.\ $A \notindep C \given B$) \cite{pearl2009causality}.

For more general graph structures, one should reason in terms of paths (i.e.\ chains of nodes connected by edges pointing in any direction), as they are the conduits for correlations propagated across the graph. Any path that does not contain a collider is said to be \emph{unblocked} or \emph{open}, and implies a potential statistical association between its endpoints. Conversely, a path containing a collider is said to be \emph{blocked} or \emph{closed}, and does not carry any indirect causal influence between its endpoints \textit{a priori} \cite{greenland2011diagrams}. If there are no unblocked paths between two variables, we conclude they are independent.
As mentioned above, however, conditioning on a collider (or on a descendant of one) may unblock previously blocked paths.


In causality, what is usually referred to as `bias' is any spurious correlation between two variables, contributed by unblocked paths beside the relationship of interest.
The `classic' prototypical configurations inducing such biases are confounding (unadjusted common cause; cf.\ Simpson's paradox \cite{pearl2009causality}) and collider bias (conditioning on a common effect; cf.\ selection bias, Berkson's paradox \cite{pearl2009causality}), and are widely studied in statistical literature \cite{hernan2004structural}.
This article in specific focused on how dataset shift results from unblocked paths between domain indicators and relevant variables, and on the consequences of (implicitly) conditioning on selection variables.
For example, in a multi-site study wherein age distributions vary across sites, it would be useful to use age alongside the image as inputs to the predictive model, to block the `site $\causes$ age $\causes$ image' path causing population shift.
This is what is normally meant in the context of predictive modelling, as in statistics and causal inference, by `adjusting/controlling for' or `conditioning on' a variable.